\documentclass[aps,twoside,superscriptaddress,twocolumn,floatfix,a4paper,prl,longbibliography]{revtex4-1}
\usepackage{color}
\usepackage{graphicx}
\usepackage[utf8x]{inputenc}
\usepackage[T1]{fontenc}
\usepackage{siunitx}
\usepackage{amstext}
\usepackage{textgreek}
\usepackage{hyperref}
\usepackage{epstopdf}
\usepackage{amsfonts}
\usepackage{amsmath}
\usepackage{multirow}

\begin{document}

\title{Experimental measurement of Hilbert-Schmidt distance between two-qubit states as means for speeding-up machine learning}

\author{Vojtěch Tr\'{a}vni\v{c}ek}
\email{vojtech.travnicek@upol.cz}
\affiliation{RCPTM, Joint Laboratory of Optics of Palacký University and Institute of Physics of Czech Academy of Sciences, 17. listopadu 12, 771 46 Olomouc, Czech Republic}

\author{Karol Bartkiewicz} \email{bark@amu.edu.pl}
\affiliation{RCPTM, Joint Laboratory of Optics of Palacký University and Institute of Physics of Czech Academy of Sciences, 17. listopadu 12, 771 46 Olomouc, Czech Republic}
\affiliation{Faculty of Physics, Adam Mickiewicz University,
PL-61-614 Pozna\'n, Poland}

\author{Antonín \v{C}ernoch} \email{acernoch@fzu.cz}
\affiliation{Institute of Physics of the Czech Academy of Sciences, Joint Laboratory of Optics of PU and IP AS CR, 17. listopadu 50A, 772 07 Olomouc, Czech Republic}
   
\author{Karel Lemr}
\email{k.lemr@upol.cz}
\affiliation{RCPTM, Joint Laboratory of Optics of Palacký University and Institute of Physics of Czech Academy of Sciences, 17. listopadu 12, 771 46 Olomouc, Czech Republic}   

\begin{abstract}
We report on experimental measurement of the Hilbert-Schmidt distance between two two-qubit states by many-particle interference. We demonstrate that our three-step method for measuring distances in Hilbert space is far less complex than reconstructing density matrices and that it can be applied in quantum-enhanced machine learning to reduce the complexity of calculating Euclidean distances between multidimensional points, which can be especially interesting for near term quantum technologies and quantum artificial intelligence research. Our results are also a novel example of applying mixed states in quantum information processing. Usually working with mixed states is undesired, but here it gives the possibility of encoding extra information as coherence between given two dimensions of the density matrix.  
\end{abstract}

\date{\today}

\maketitle
\paragraph*{Introduction.}
Quantum information protocols such as teleportation \cite{marcikic200450km, bib:Ma:teleport143} and cryptography \cite{bib:bb84,bib:e91, bib:R05} established in the field of quantum information processing \cite{bib:alber:quantum_inform, bib:bowmeester:quantum_inform} have a significant impact on modern communications \cite{bib:bb84:exper, ralph99CVcrypto, bib:croal:freespace}. In fact, early quantum communications networks based on quantum teleportation have already been reported \cite{bib:pirandola:tele:nature, bib:loock:CVtele, bib:castelvecchi} and experimentally realized \cite{bib:yonezawa:nature}. Their physically guaranteed security \cite{zurek1982single} and potential for scalability makes them a preferable choice for future communications networks. In quantum communications the quality of a transmission channel is crucial. It is due to security reasons, where imperfections of the communication channel lead to signal degradation known as noise.  This noise can be subsequently exploited by a potential eavesdroppers \cite{bib:bartkiewicz:eaves,Bartkiewicz2017}. Therefore, tools for diagnostics of the transmission channels are in demand. In quantum communications theory one can quantify the accuracy of a signal transmission by measuring the distance in Hilbert space between the transmitted and received states. The most prominent distance measures include Uhlmann-Jozsa fidelity (Bures metrics), trace distance, and \emph{Hilbert-Schmidt distance} (HSD) (for overviews see, e.g., \cite{bengtsson2017geometry,PhysRevA.84.032120,PhysRevA.99.032336}). 

%====================================
\begin{figure}
\includegraphics[width=8.5cm]{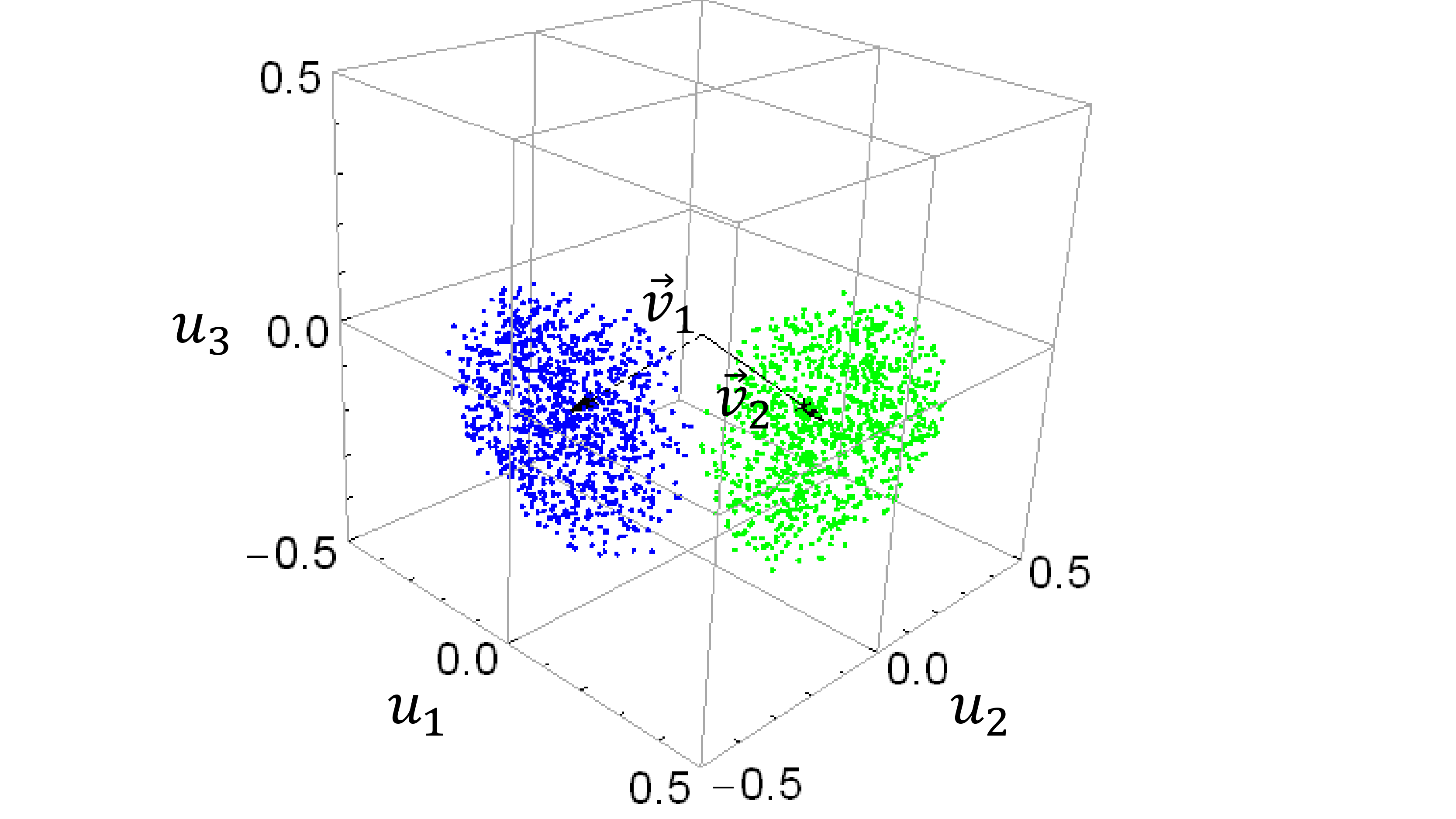}
\caption{\label{fig:clusters} An example of two clusters of 1000 3D points assigned to the nearest center of a cluster (given by arrows $\vec{v}_1$ or $\vec{v}_2$). By encoding vectors $\vec{u}$ and $\vec{v}_{i}$ for $i=1,2$ as density matrices of qubits and by measuring distances between them as HSD we properly assign all the points to one of the clusters. Thus, for two-qubit states can classify 15-dimensional points. Note that all the points are embedded in a Bloch ball of radius $\frac{1}{2}.$}
\end{figure}
%====================================

These distance measures are also essential for a field of quantum machine learning. Where a common method for a classification algorithms (e.g. \textit{k}-means) is to perform a distance measurement among $M$ sample vectors of dimension $N$. This procedure is a core subroutine for other machine learning algorithms, e.g., supervised and unsupervised nearest-neighbor algorithm. It has been already demonstrated that by using quantum resources one can reduce the complexity of the algorithm from $\mathcal{O}[\mathrm{poly}(MN)]$  to  $\mathcal{O}[\log(MN)]$ \cite{Lloyd13,PhysRevLett.114.110504,Biamonte2017Nature}. Here, we demonstrate that by measuring the distance in terms of HSD we obtain the complexity of the distance-measuring algorithm to $\mathcal{O}(\log{N})$ by using a different approach than in Ref.~\cite{Lloyd13}. The HSD is defined as
\begin{equation}
D_{HS}(\hat\rho_1,\hat\rho_2) \equiv \sqrt{\mathrm{Tr}[(\hat\rho_{1}-\hat\rho_{2})^2]},
\label{eq:HSD}
\end{equation}
where $\hat\rho_1$ and $\hat\rho_2$ are the density matrices representing the two quantum in general mixed states. 
The HSD is a Riemannian metrics, which makes it appropriate for applying in machine learning problems. Moreover, in contrast to trace distance, HSD is non-increasing under decoherence \cite{bengtsson2017geometry,PhysRevA.84.032120}. For simplicity, let us explain how to implement the \textit{k}-means algorithm for finding 2 clusters of 3D points enclosed in a cube using qubits.  A density  matrix for a qubit can be expressed via Pauli matrices $\hat\sigma_i$ and the identity matrix $\hat I$ as $\hat\rho =\tfrac{1}{2}\hat{I} + u_1\hat\sigma_x + u_2\hat\sigma_y + u_3\hat\sigma_z$. Let use this kind of matrix to encode a data point $\vec{u} = (u_1,u_2,u_3)$ for $|\vec{u}|\leq \frac{1}{2}$ and $i=1,2,3.$ The task is to assign different $N$-dimensional data points (here $N=3$) to $k$-clusters (let us set $2$ clusters) with sample reference vectors $\vec{v}_1$ and $\vec{v}_2.$ A data point $\vec{u}$ is classified to the closest cluster. It turns that by a proper choice of mapping between vectors and density matrices, we can ensure that Euclidean distance $|\vec{v}_{1,2} -\vec{u}|$ and HSD are equal up to a constant factor. Thus, by assigning data points to a cluster corresponding to the nearest reference vector, depending on the distribution of $\vec{u},$ we can end up with two clearly separable clusters as shown in Fig.~\ref{fig:clusters}. In the next step of the \textit{k}-means algorithm, new positions of centers of clusters are found as mean positions of points belonging to a given cluster. The classification process is repeated. If points were not change their assigned clusters, the algorithm is terminated.  Note that the same applies to larger systems, e.g., for a 15-dimensional (in Hilbet-Schmidt space) physically-accessible Bloch ball (or the inscribed hypercube if the components of $\vec{u}$ need to represent data from a segment $[-l,l],$ where $l$ is the size of the hypercube) the corresponding state is given as $\hat\rho =\tfrac{1}{4}\hat I\otimes \hat I + \sum_{i=1,2,3}(u_i\hat\sigma_i\otimes \hat I + u_{i+3} \hat I\otimes\hat \sigma_i + \sum_{j=1,2,3}u_{j+3(i+1)}\hat\sigma_i\otimes\hat\sigma_j),$ where $|\vec{u}| \leq \sqrt{\frac{3}{8}}.$  For $D$-dimensional Hilbert space a density matrix contains $(D^2-1)$ independent parameters given as vector $\vec{u}$ and as many generalized Pauli operators [i.e, traceless generators of $SU(D)$], where $|\vec{u}|\leq \sqrt{\frac{D-1}{2D}}$. This fact makes the complete quantum state tomography a very challenging problem as it requires an exponentially large number of measurements in relation to the number of qubits constituing the composite system (see, e.g., \cite{ParisBook2004,PhysRevA.90.062123,srep19610}). However, this otherwise problematic feature also opens a new possibility to encode $N=D^2-1$ parameters in a $D$-dimensional density matrix (i.e., Hilbert-Schmidt space). Once this is done for $M$ states a constant number of times, each distance can be measured in only $3$ steps. This is because the HSD can be expressed by first-order overlaps $O(\hat\rho_i, \hat\rho_j)$ as described in Ref.~\cite{PhysRevA.99.032336,ZhangPRA2017,FilipPRA2002}
\begin{equation}
D_{HS}(\hat\rho_1,\hat\rho_2) = \sqrt{O(\hat\rho_1,\hat\rho_1)+O(\hat\rho_2,\hat\rho_2)-2O(\hat\rho_1,\hat\rho_2)},
\end{equation}
where the there directly measured observables are defined as
$O(\hat\rho_i,\hat\rho_j) = \mathrm{Tr}(\hat\rho_i\hat\rho_j).$
If $\hat\rho_1 = \hat\rho_2,$ we measure purity as discussed, e.g., in Refs.~\cite{bib:bovino:purity,PhysRevA.88.052104,PhysRevA.94.052334}. Each overlap or other functions of overlaps can be measured directly by utilizing multi-particle interactions between copies of the investigated states~\cite{EkertPRL2002,AlvesPRL2004,
Brun2004,JeongJOSAB2004,PhysRevA.88.052104,PhysRevA.95.030102,IslamNature2015,ZhangPRA2017,PhysRevA.99.032336}. In contrast by applying full quantum tomography (see, e.g., \cite{ParisBook2004}) $(D^2-1)$ measurements are required  to calculate the value of HSD. For technical reasons we measure each overlap by utilizing $4$ \emph{positive-valued measures} (POVMs). For $D=4$ this amounts to $12$ POVMs for obtaining a single value of $D_{HS}$ in contrast to $32$ measurements needed in case of applying quantum state tomography ($16$ if two copies of a sate are used in parallel). This discrepancy can become even grater in case of larger values of $D.$ The number of the required observables to measure in each of the $3$ steps for a multi-qubit overlap depends linearly on the number of qubits forming the density matrix, i.e., $n=\log_2(D)$ as $\mathcal{O}(n)=\mathcal{O}[\log_2(\sqrt{N+1})]$ (see, e.g., Refs. \cite{JeongJOSAB2004,AlvesPRL2004,IslamNature2015,ZhangPRA2017}). Thus, the complexity of the distance measurement is $\mathcal{O}(\log {N})$.

%====================================
\begin{figure}
\begin{center}
\includegraphics[width=7cm]{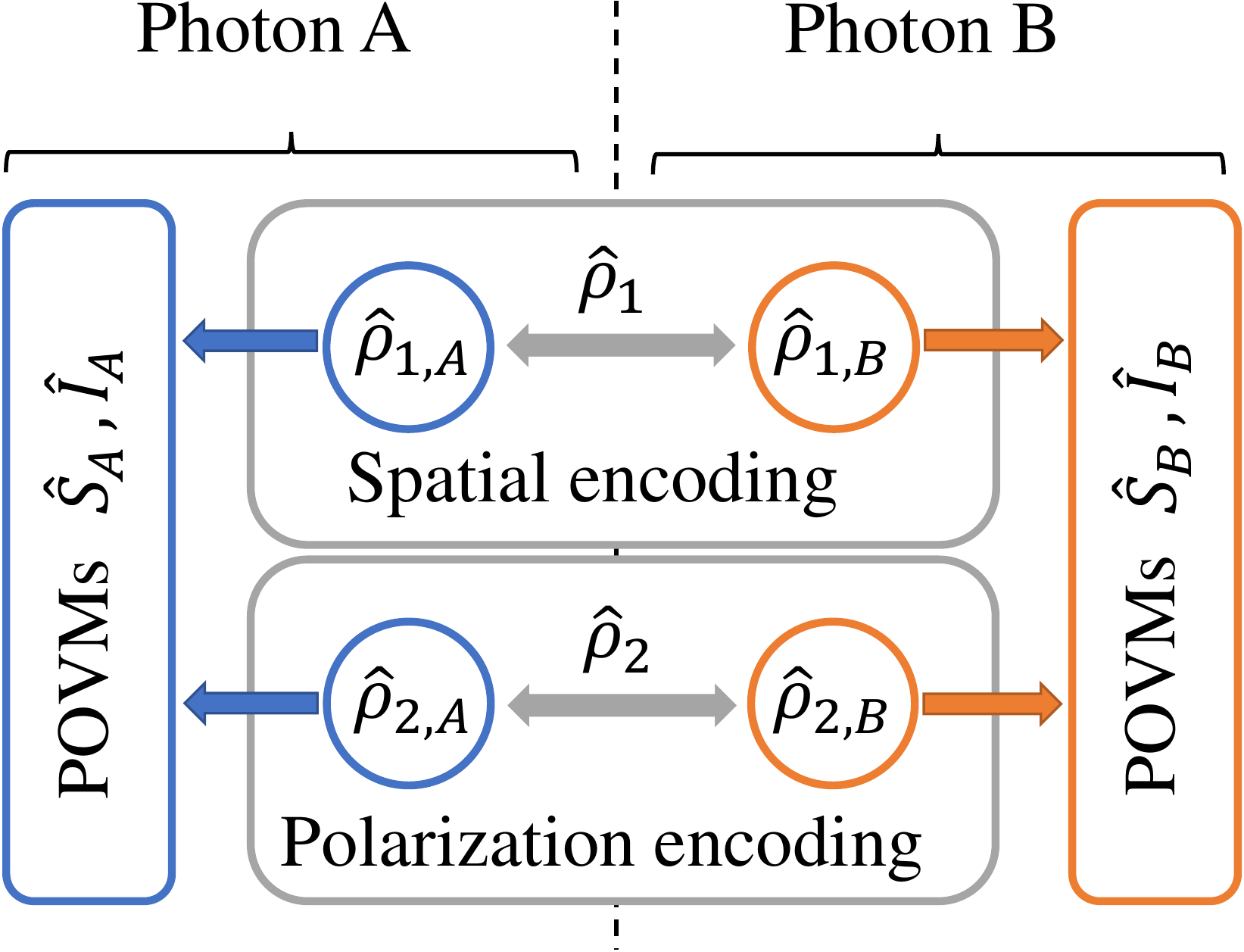}
\end{center}
\vspace*{-5mm}
\caption{\label{fig:concept} (color online) Conceptual scheme for measuring the Hilbert-Schmidt distance between two-qubit states. In general two different states $\hat{\rho}_1$ and $\hat{\rho}_2$ are encoded into polarization and spatial modes of photon $A$ and $B$ respectively. Photons $A$ and $B$ are then simultaneously measured by POVMs $\hat{I}$ and $\hat{\mathrm{S}},$ where the two degrees of freedom are addressed holistically at the same time. The operators $\hat{I}$ and $\hat{\mathrm{S}}$ are the identity and singlet state projection where $\hat{\mathrm{S}} = |\Psi^-\rangle\langle\Psi^-|$.}   
\end{figure}

\begin{figure}
\begin{center}
\includegraphics[width=7cm]{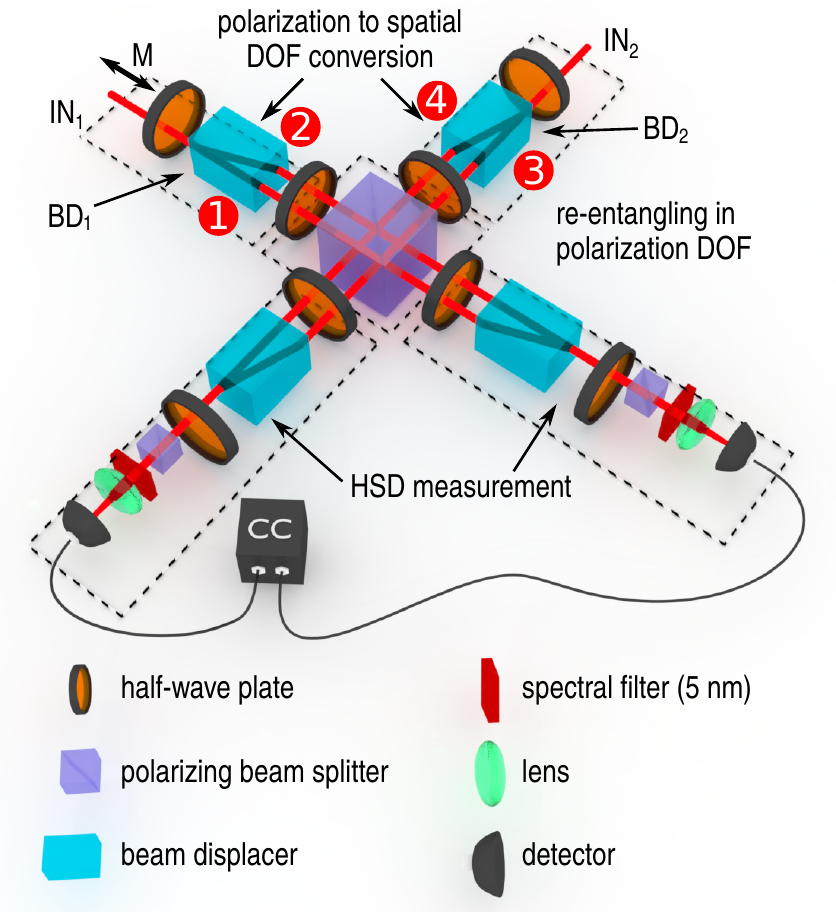}
\end{center}
\vspace*{-5mm}

\caption{\label{fig:setup} (color online) Experimental setup for measuring Hilbert-Schmidt distance of photonic two-qubit states. Spatial modes are labeled by numbers 1--4.}   
\end{figure}

\begin{figure*}
\begin{center}
\includegraphics[width=0.93\textwidth]{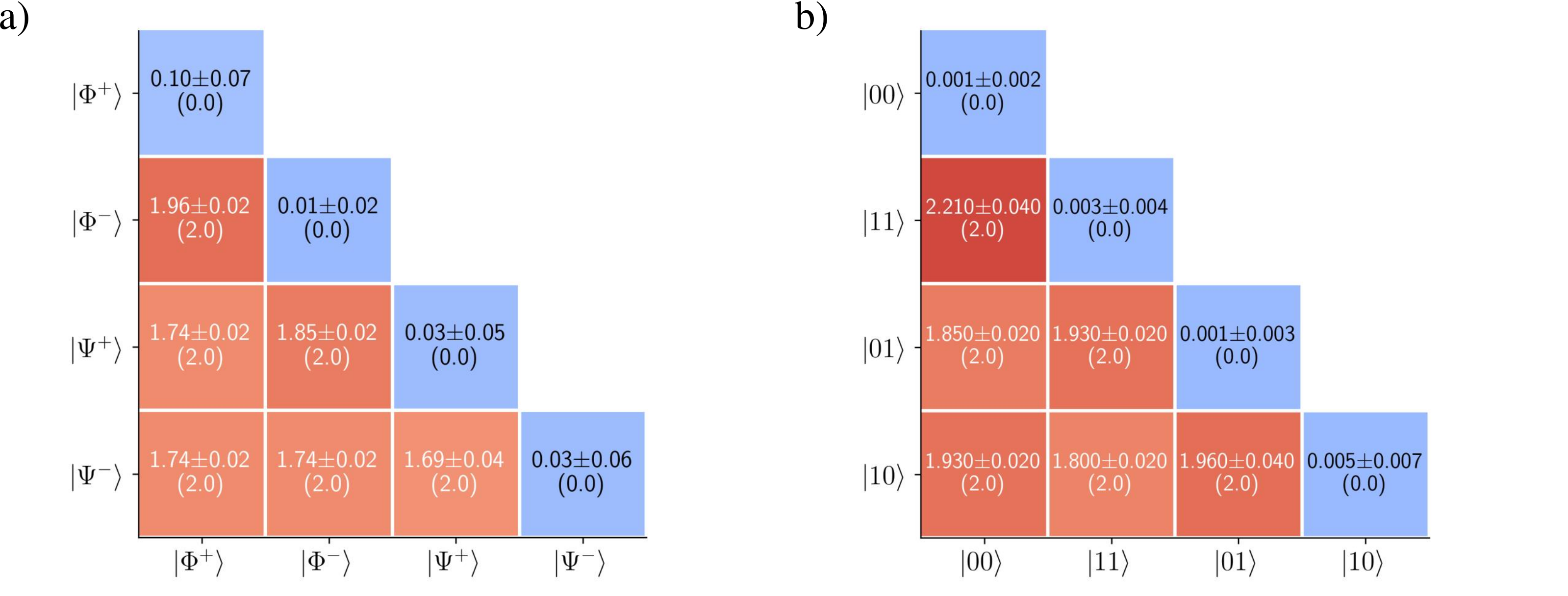}  
\end{center}
\vspace*{-5mm}
\caption{\label{fig:EE}(color online) Experimental results and theoretical values (in parentheses) of the second power of Hilbert-Schmidt distance $D_{HS}^2$ between: (a) Bell states, (b) separable states. The vertical and horizontal axes represent polarization and spatial encoding respectively (see Fig.~\ref{fig:concept}).}    
\end{figure*}
%====================================

\paragraph*{Experimental setup.}
Let us demonstrate the measurement of HSD with an linear-optical experiment with photons as information carriers. Here, the HSD is measured for two-qubit states by simultaneous interaction between 4 qubits. A usual approach uses 4 photons of only one degree of freedom (DOF) such as polarization~(see, e.g., Ref.~\cite{ZhangPRA2017}). Here we utilize two DOFs (polarization and spatial) to encode two qubits, therefore, only two photons were needed. This way one achieves much higher detection rates which makes the experiment considerably faster. A horizontally polarized photon (i.e., subsystem $X=A,B$) encodes the logical state $|0\rangle_{Xp}$, a vertically-polarized photon state $|1\rangle_{Xp}$. Similarly, its spatial modes $1$ and $3$ encode logical state $|0\rangle_{As}$ and $|0\rangle_{Bs}$, modes $2$ and $4$ logical state $|1\rangle_{As}$ and $|1\rangle_{Bs}$ (see scheme in Fig.~\ref{fig:setup}).

The two photons are generated in a crystal cascade (known as the Kwiat source \cite{PhysRevA.60.R773}) pumped by pulsed Paladine (Coherent) laser at $\lambda$ = \SI{355}{\nano\meter} with \SI{200}{\milli\watt} of mean optical power and repetition rate of \SI{120}{\mega\hertz}. The source consists of two BBO ($\beta$-BaB$_{2}$O$_{4}$) crystals and generates polarization--entangled photon pairs at $\lambda$ = \SI{710}{\nano\meter}, i.e.,  $|\Psi\rangle = \mathrm{cos}(\alpha)|HH\rangle + e^{i\theta}\mathrm{sin}(\alpha)|VV\rangle.$
In this state, $H$ and $V$ stand for horizontal and vertical polarizations. The rates and mutual phase shift between horizontally and vertically polarized photons can be tuned by adjusting the pump beam polarization or by tilting one of the beam displacers ($BD_1$ or $BD_2$ in Fig.~\ref{fig:setup}). By doing so one can prepare states with various amount of entanglement. Each photon from the generated pair is coupled into a single--mode optical fiber and brought to one input port of the experimental setup depicted in Fig.~\ref{fig:setup}.
The photons then pass through beam displacers where the initial polarization encoding is transformed into spatial encoding. Afterwards the photons interact on the polarizing beam splitter (PBS) where a second, in principle different, quantum state is encoded into polarization DOF. As a result, two, in principle different, two-qubit states are encoded into the two DOFs.
The two states are then subjected to projective measurements as discussed below and accompanied by post-selection. The photons are filtered by \SI{5}{\nano\meter} interference filters, coupled into single--mode optical fibers and brought to single-photon detectors. Motorized translation M ensures temporal overlap of the photons on PBS. To demonstrate versatility of this approach, we have measured the HSD between $4$ Bell states, 4 separable states, Werner states, and between Werner and Horodecki states. 

To measure the HSD between any two states ($\rho_1,\rho_2$) the first-order overlap has to be measured in three configurations, i.e., $O(\rho_{1},\rho_{1})$, $O(\rho_{2},\rho_{2})$ and $O(\rho_{1},\rho_{2})$. The first two configurations correspond to the situation when $\rho_1$ ($\rho_2$) is encoded into both DOFs. During the last configuration $\rho_1$ and $\rho_2$ are encoded each in one DOF. Measurement of each first-order overlap $O(\rho_1,\rho_2)$ is split into a measurement of 4 POVMs on each photon across its DOFs, i.e., $\hat{I}_A\otimes\hat{I}_B,$ $\hat{S}_A\otimes\hat{I}_B,$  $\hat{I}_A\otimes\hat{S}_B,$ and $\hat{S}_A\otimes\hat{S}_B.$ The POVMs are identity ($\hat{I}$) and singlet ($\hat{S}$) projections that were implemented by suitable rotation of HWPs behind the PBS. For example the POVM $\hat{I}_A\otimes\hat{I}_B$ consists of all combinations of local projections, i.e., $|H_{1},H_{3}\rangle_{A,B}, |H_{2},H_{3}\rangle_{A,B}, ..., |V_{2},V_{4}\rangle_{A,B}$, while the $\hat{S}_A\otimes\hat{S}_B$ consists of projections $\frac{1}{\sqrt{2}} (|H_{4}\rangle-|V_{3}\rangle)_B$ and $\frac{1}{\sqrt{2}}(|H_{2}\rangle-|V_{1}\rangle)_A$. Both these POVMs can be implemented in a single step, but in this experiment they were implemented as a series of Von-Neumann projections. The coincidence rates corresponding to specific POVMs are labeled $f_{\hat{x}\hat{y}}$, where $\hat{x}, \hat{y} \in \{\hat{I}, \hat{S}\},$ where $\hat{x}$ and $\hat{y}$ are associated with photon A and B, respectively. These values are obtained by summing up the coincidence rates associated with respective Von-Neumann projections. The mean value of the overlap operators relates to these rates as
\begin{equation}
O(\rho_{1},\rho_{2}) = 1-2(f_{\hat{S} \hat{I}} +f_{\hat{I} \hat{S}} -2f_{\hat{S} \hat{S}})/f_{\hat{I} \hat{I}}.
\end{equation}
Note that POVMs associated with $f_{\hat{I} \hat{I}}$  measures photon rate and is needed for normalization. In case of a stable photon source and know setup parameters this value is constant and state-independent. The same is true for POVMs $\hat{I}_A$ and $\hat{I}_B$ separately.

\paragraph*{Experimental results.}
First, we have measured the distances between $4$ Bell states  $|\Phi^{\pm}\rangle  = \frac{1}{\sqrt{2}}(|00\rangle\pm|11\rangle)$ and $|\Psi^{\pm}\rangle = \frac{1}{\sqrt{2}}(|01\rangle\pm|10\rangle)$. Encoding of the states into the DOFs was implemented by a suitable choice of pump beam polarization, rotation of the HWPs and by tilting one of the beam displacers ($BD_1$). We have decided to plot second power of the HSD denoted $D_{HS}^2$ so it is linear in terms of the physically measured quantities. The obtained experimental and theoretically calculated values of the second power of HSD between Bell states are  shown in Fig.~\ref{fig:EE}a. Next, we have measured the HSD between separable states $|00\rangle$, $|11\rangle$, $|01\rangle$ and $|10\rangle$ and visualized the obtained values of $D_{HS}^2$ in Fig.~\ref{fig:EE}b. 

%====================================
\begin{figure*}
\begin{center}
\includegraphics[width=0.93\textwidth]{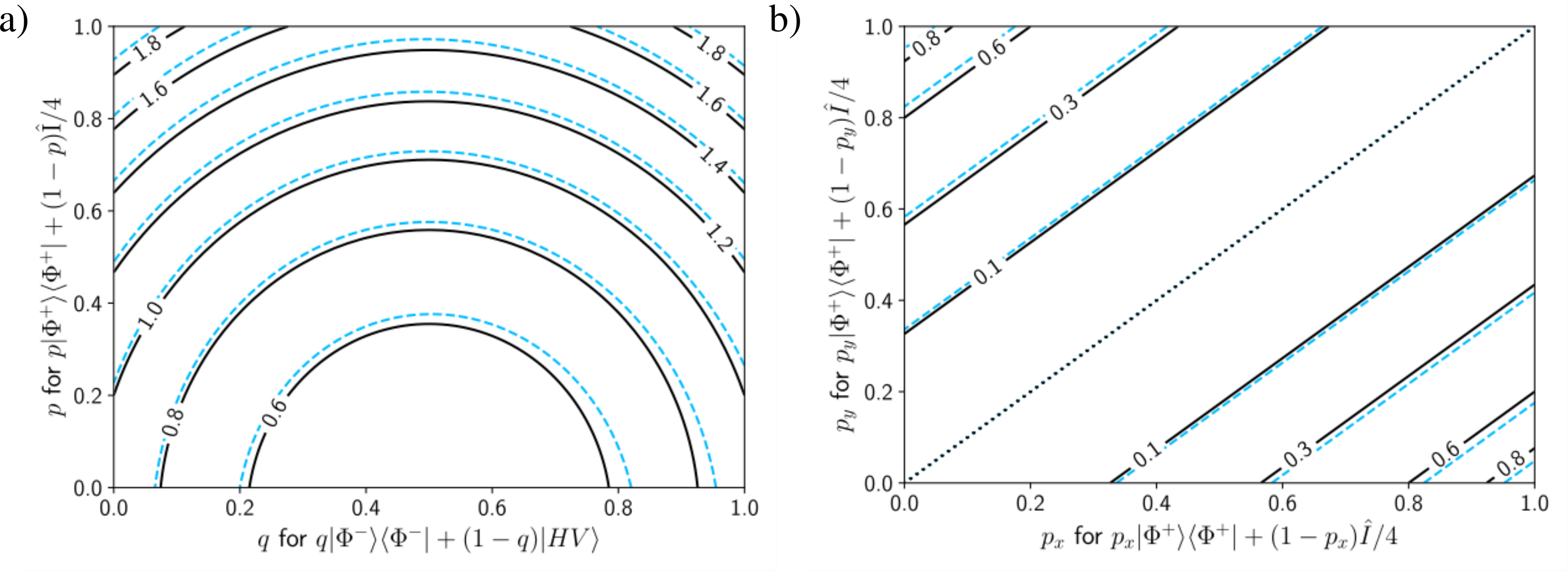}
\end{center}
\vspace*{-5mm}
\caption{\label{fig:WW} (color online) Experimentally obtained values of $D_{HS}^2$ (a) between two Werner states  and (b) between Werner and Horodecki states for various weight parameters $(p_x,p_y)$ or $(p,q)$ are represented by a corresponding light-shaded contours slightly shifted with respect to the labeled black contours representing the theoretical values. The vertical and horizontal axes represent polarization and spatial encoding respectively (see Fig.~\ref{fig:concept}).}        
\end{figure*}
%====================================

In the third part of the experiment, we have calculated the values of $D_{HS}^2$ between Werner states which up to a local unitary transformation can be expressed in a form of a weighted sum of maximally entangled and maximally mixed state
\begin{equation}
\hat{\rho}_W = p|\Phi^+\rangle\langle\Phi^+| + \tfrac{1}{4}(1-p)\hat{I}.
\end{equation}
In case of the mixed state, the outcome of each Von-Neumann projection was obtained by accumulating coincidence rates associated with 4 Bell states, i.e., making use of the identity
$\hat{\rho}_1 \otimes \hat{\rho}_2 = \frac{1}{4}(|\Psi^+\rangle\langle\Psi^+|+|\Psi^-\rangle\langle\Psi^-|+ |\Phi^+\rangle\langle\Phi^+|+|\Phi^-\rangle\langle\Phi^-|)= \frac{1}{4} I\otimes I.$ Subsequently, we have calculated the $D_{HS}^2$ between Werner states for various values of the weight parameter $p$. The results are visualized in Fig.~\ref{fig:WW}a.
Finally, we have calculated the $D_{HS}^2$ between Werner and Horodecki states. Horodecki states can be expressed in form of a weighted sum of maximally entangled and separable state
\begin{equation}
\hat{\rho}_H = q|\Phi^-\rangle\langle\Phi^-| + (1-q)|HV\rangle.
\end{equation}
Therefore, we had to measure the overlap between states $|\Phi^+\rangle$ ($|\Phi^-\rangle$) and $|01\rangle$ encoded in polarization and spatial mode respectively. Rest of the necessary overlaps were calculated in the same way as explained above. The values of $D_{HS}^2$ between Werner and Horodecki states for various weight parameters $p$ and $q$ are visualized in Fig.~\ref{fig:WW}b.

\paragraph*{Conclusions.}
We have reported on experimental measurement of the Hilbert-Schmidt distance between two-qubit states by the method of many-particle interference. This method allows to measure HSD between two two-qubit density matrices by performing $3$ overlap measurements ($4$ POVMs per overlap) instead of $32$ measurements required to reconstruct two two-qubit mixed states. The obtained results are in good agreement with theoretical predictions. To demonstrate the versatility of our approach we measured HSD between assorted two-qubit states. The HSDs between identical Bell states are sufficiently close to theoretical values. On the other hand, distances between orthogonal Bell states do not deviate from theoretical values by more then 15\%. This error is partially caused by the linearization of the Eq~(\ref{eq:HSD}). We have obtained similar results for the separable states, however, the deviation from theoretical prediction is not as high due to lower complexity of the states. We have also interpolated the HSDs between Werner states and between Werner and Horodecki states for various values of the weight parameters. The results are in good agreement with theoretical values represented by the contours in Fig.~\ref{fig:WW}. We believe that these results can motivate subsequent research on the topic of quantum channel characterization and quantum machine learning. Especially in the latter, measuring distances between multidimensional points efficiently can reduce the computational complexity of supervised and unsupervised machine learning. Thus, our results can be inspiring for near term quantum technologies which would exhibit speedup in comparison to the best currently known classical solutions. Our results are also a novel example of applying mixed states for quantum information processing. Usually working with mixed states is undesired, but here it gives the possibility of encoding extra information in coherence between given two dimensions of density matrix. 
\paragraph*{Acknowledgements.}
\begin{acknowledgments}
Authors thank Cesnet for providing data management services. Authors acknowledge
financial support by the Czech Science Foundation under the project No. 19-19002S. The authors also acknowledge the projects Nos. LO1305 and CZ.02.1.01./0.0/0.0/16\textunderscore 019/0000754 of the Ministry of Education, Youth and Sports of the Czech Republic financing the infrastructure of their workplace and VT also acknowledges the Palacky University internal grant No. IGA-PrF-2019-008.
\end{acknowledgments}

%\bibliography{gacr}{}

\end{document}